# Unique mechanisms from finite two-state trajectories

Ophir Flomenbom & Robert J. Silbey

*Department of Chemistry, Massachusetts Institute of Technology, Cambridge, MA 02139*

**Single molecule data made of *on* and *off* events are ubiquitous. Famous examples include enzyme turnover, probed via fluorescence, and opening and closing of ion-channel, probed via the flux of ions. The data reflects the dynamics in the underlying multi-substate *on-off* kinetic scheme (KS) of the process, but the determination of the underlying KS is difficult, and sometimes even impossible, due to the loss of information in the mapping of the mutli-dimensional KS onto two dimensions. A way to deal with this problem considers canonical (unique) forms. (Unique canonical form is constructed from an infinitely long trajectory, but many KSs.) Here we introduce canonical forms of reduced dimensions that can handle any KS (i.e. also KSs with symmetry and irreversible transitions). We give the mapping of KSs into reduced dimensions forms, which is based on topology of KSs, and the tools for extracting the reduced dimensions form from finite data. The canonical forms of reduced dimensions constitute a powerful tool in discriminating between KSs.**

*I - Introduction*

Single molecule data sets are turned, in many cases, into trajectories of *on* and *off* periods (waiting times), Fig. 1A. Examples include the passage of ions and biopolymers through individual channels [3-5], activity and conformational changes of biopolymers [1-2, 6-16], diffusion of molecules [17-20], and blinking of nano-crystals [21-24]. The mechanism of the observed process is usually described by a multi-substate *on-off* Markovian kinetic scheme (KS) [25-39], Fig. 1B; see Refs. 40-55 for related descriptions. The KS describes a discrete conformational energy landscape of a biomolecule, chemical kinetics with (or without) conformational changes, or environmental changes, stands for quantum states, etc. The underlying stochastic dynamics of the process in the multi-substate *on-off* KS is thus encoded in the two-state trajectory (the stochastic signal changes value only when transitions between substates of different states in the KS take place). From single molecule experiments, we wish to learn as much as possible about the underlying KS, much more than the general properties obtained from bulk measurements.

However, determining the KS from the two-state trajectory is difficult since the number of the substates in each of the states, $L_x$ ($x =$ *on, off*), is usually large, and the connectivity among the substates is usually complex. A more fundamental difficulty in finding the correct KS arises from the equivalent of KSs, i.e. there are KSs that lead to the same two-state trajectories in a statistical sense [28-30, 32-36]. A way to deal with this issue is to use canonical (unique) forms rather than KSs [32-35]: the space of KSs is mapped into a space of canonical forms. A given KS is mapped to a unique canonical form but many KSs can be mapped to the same canonical form. Underlying

KSs with the same canonical form are equivalent to each other, and cannot be discriminated based on the information in a single infinitely long two-state trajectory.

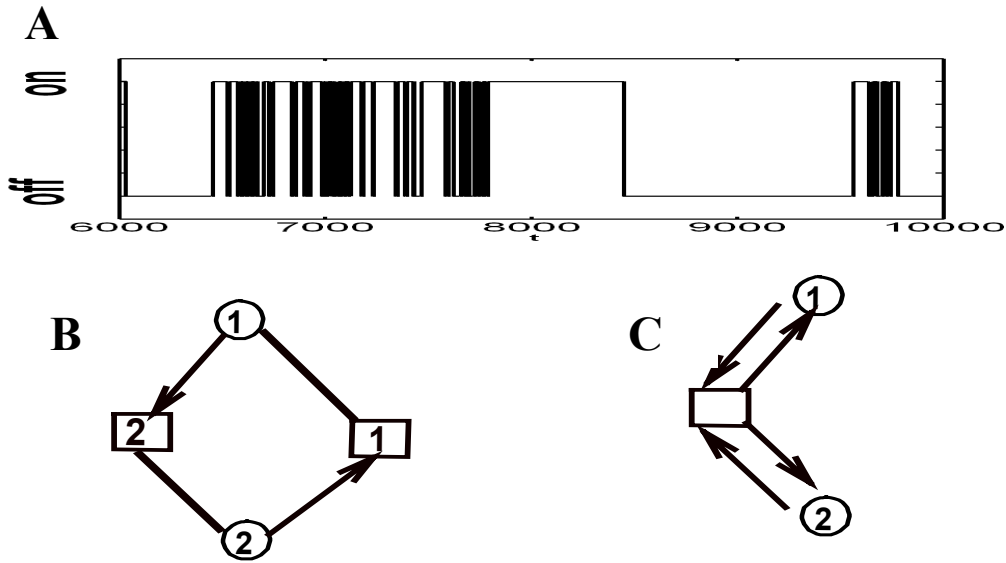

**FIG 1** A two-state trajectory (**A**), the KS (**B**), and the RD form (**C**). The KS **B** has $L_{on}=2$ (squared substates), $L_{off}=2$ (circled substates), irreversible transitions, and $N_{on}=M_{on}=2$ and $N_{off}=M_{off}=2$. For generating the data, we take the following transition rate values ($k_{ji}$ connects substates $i \to j$): $k_{1_{off}1_{on}} = 0.3$, $k_{2_{off}2_{on}} = 0.02$, $k_{1_{on}1_{off}} = 0.425$, $k_{2_{on}1_{off}} = 0.075$, $k_{1_{on}2_{off}} = 0.0085$, $k_{2_{on}2_{off}} = 0.0015$ (with arbitrarily units). The equality of the ratios, $k_{2_{on}2_{off}}/k_{1_{on}2_{off}} = k_{2_{on}1_{off}}/k_{1_{on}1_{off}} \equiv p_R/p_L$ ($p_L + p_R = 1$), means that the KS is symmetric in the sense that the ranks of the 2D WT-PDFs of successive $x, y$ (=*on, off*) events are all equal one, except $R_{on,off}$ which equal two. As a result, the RD form (**C**) has one *on* substate and two *off* substates. The RD form has also direction dependent WT-PDFs for the *on* to *off* connections, $\varphi_{on,11}(t) = p_L k_{1_{off}1_{on}} e^{-k_{1_{off}1_{on}}t}$, $\varphi_{on,21}(t) = p_R k_{2_{off}1_{on}} e^{-k_{2_{off}1_{on}}t}$, and $\varphi_{off,11} = k_{1_{off}} e^{-k_{1_{off}}t}$, $\varphi_{off,12}(t) = k_{2_{off}} e^{-k_{2_{off}}t}$. Here, $k_i = \sum_j k_{ji}$, and $\varphi_{x,ij}(t)$ is the WT-PDF that connects substates $x_j \to y_i$ in the RD form.

Here, we give canonical forms of reduced dimensions (RD) [33], e.g. Fig. 1C. RD forms have many advantageous over other canonical forms that were previously suggested. RD forms can handle *any* KS, i.e. KSs with irreversible connections and/or symmetry (symmetry means that the spectrum of the waiting time probability density function (WT-PDF) for the single *x* (=*on, off*) periods is degenerate), and constitute a powerful tool in discriminating among KSs, because the mapping of a KS into a RD form is based on the KS's *on-off* connectivity and therefore can be done, to a large extent, without actual calculations. Using this property, we give an ensemble of relationships between properties of the data, the topology of the RD forms, and properties of KSs. These relationships are useful in discriminating KSs and in the analysis of the data. With respect of the actual analysis of the data, we give a simple procedure for finding the RD form from finite data. It turns out that RD forms can be constructed from data sets fairly accurately, and, importantly, much more accurately than other mechanism (at least in the cases studied by us, which represent commonly encountered mechanisms from two-state trajectories.)

This paper is laid out as follows. Section II gives the mathematical formulations of the system in terms of the master equation and the path representation. Section III introduces the canonical forms of reduced dimensions. RD forms stem from the path representation. Section IV builds a RD form from finite data, and section V concludes.

## II - *Mathematical formulations*

In this section, we express the WT-PDFs for single periods, $\phi_x(t)$, *x = on, off*, and for joint successive periods, $\phi_{x,y}(t_1,t_2)$, *x, y = on, off*, in terms of both the master equation and the path representation. The relationship between the two representations is made. *On-off* KSs are commonly described in terms of the master equation, but our canonical forms are naturally related to the path representation. This will be shown in the next section.

***II.1. Matrix formulation of the system*** The problem at hand is formulated in terms of a (single) random walker in an *on-off* KS. The time-dependent occupancy probabilities of state *x* for the coupled *on-off* process, $\vec{P}_x(t)$, with the elements $(\vec{P}_x(t))_i = P_{x,i}(t)$ for $i = 1,...,L_x$ ($L_x$ is the number of substates in state *x*), obey the reversible master equation,

$$\frac{\partial}{\partial t}\begin{pmatrix} \vec{P}_{on}(t) \\ \vec{P}_{off}(t) \end{pmatrix} = \begin{pmatrix} \mathbf{K}_{on} & \mathbf{V}_{off} \\ \mathbf{V}_{on} & \mathbf{K}_{off} \end{pmatrix} \begin{pmatrix} \vec{P}_{on}(t) \\ \vec{P}_{off}(t) \end{pmatrix}. \tag{1}$$

In Eq. (1), matrix $\mathbf{K}_x$, with dimensions $[\mathbf{K}_x] = L_x, L_x$, contains transition rates among substates in state *x*, and 'irreversible' transition rates from substates in state *x* to substates in state *y*. (The 'irreversible' transition rates are given on the diagonal, and are called irreversible because matrix $\mathbf{K}_x$ does not contain the back transition rates from state *y* to state *x*). Matrix $\mathbf{V}_x$, with dimensions $[\mathbf{V}_x] = L_y, L_x$, contains transition rates between states $x \to y$, where $(\mathbf{V}_x)_{ji}$ is the transition rate between substates $i_x \to j_y$. $\vec{P}_x(ss)$ is the vector of occupancy probabilities in state *x* at steady state ($t \to \infty$). $\vec{P}_x(ss)$ is found from Eq. (1) for vanishing time derivative. $\phi_x(t)$ and $\phi_{x,y}(t_1, t_2)$ are given in terms of the matrices in Eq. (1),

$$\phi_x(t) = \vec{1}_y^T \mathbf{V}_x \mathbf{G}_x(t) \mathbf{V}_y \vec{P}_y(ss) / N_x, \tag{2}$$

and,

$$\phi_{x,y}(t_1, t_2) = \vec{1}_x^T \mathbf{V}_y \mathbf{G}_y(t_2) \mathbf{V}_x \mathbf{G}_x(t_1) \mathbf{V}_y \vec{P}_y(ss) / N_x, \tag{3}$$

where $N_x = \vec{1}_x^T \mathbf{V}_y \vec{\mathbf{P}}_y(ss)$ and $\vec{1}_x^T$ is the summation row vector of $1, L_x$ dimensions, $[\vec{1}_x^T] = 1, L_x$. (The expression for $\phi_{x,x}(t_1,t_2)$ is obtained from Eq. (3) when plugging in the factor $\mathbf{V}_y \overline{\mathbf{G}}_y(0)$, $\phi_{x,x}(t_1,t_2) = \vec{1}_y^T \mathbf{V}_x \mathbf{G}_x(t_2) \mathbf{V}_y \overline{\mathbf{G}}_y(0) \mathbf{V}_x \mathbf{G}_x(t_1) \mathbf{V}_y \vec{\mathbf{P}}_y(ss) / N_x$. Here, $\overline{\mathbf{G}}_y(0)$ is the Laplace transform of $\mathbf{G}_y(t)$, $\overline{\mathbf{G}}_y(s) = \int_0^\infty \mathbf{G}_y(t) e^{-st} dt$, at $s=0$.) $\mathbf{G}_x(t)$ in Eqs. (2) - (3) is the Green's function of state $x$ for the irreversible process, $\partial \mathbf{G}_x(t) / \partial t = \mathbf{K}_x \mathbf{G}_x(t)$, with the solution,

$$\mathbf{G}_x(t) = \exp(\mathbf{K}_x t) = \mathbf{X} \exp(\lambda_x t) \mathbf{X}^{-1}. \tag{4}$$

The second equality in Eq. (4) follows from a similarity transformation $\lambda_x = \mathbf{X}^{-1} \mathbf{K}_x \mathbf{X}$, and all the matrices in Eq. (4) have dimensions $L_x, L_x$. Non symmetric KSs must have non-degenerate eigenvalue matrices $\lambda_x$s.

***II.2. Path representation of the WT-PDFs*** Our approach is based on expressing the WT-PDFs in Eqs. (2)-(3) in path representation that utilizes the *on-off* connectivity of the KS. The *on-off* process is separated into two irreversible processes that occur sequentially, and we have for $\phi_{x,y}(t_1,t_2)$ ($x \neq y$),

$$\phi_{x,y}(t_1,t_2) = \sum_{n_y=1}^{N_y} \left( \sum_{n_x=1}^{N_x} W_{n_x} f_{n_y n_x}(t_1) \right) F_{n_y}(t_2) = \sum_{m_x \in \{M_x\}} \left( \sum_{n_x=1}^{N_x} \sum_{n_y=1}^{N_y} W_{n_x} \tilde{f}_{m_x n_x}(t_1) \omega_{n_y m_x} F_{n_y}(t_2) \right). \tag{5}$$

(A sum $z_x \in \{Z_x\}$ is a sum over a particular group of $Z_x$ substates. In Appendix A expressions for $\phi_x(t)$ and $\phi_{x,x}(t_1,t_2)$ in path representation are given). Equation (5) emphasizes the role of the KS's topology in expressing the $\phi_{x,y}(t_1,t_2)$s. $N_x$ and $M_x$ are the numbers of initial and final substates in state $x$ in the KS, respectively. Namely, each event in state $x$ starts at one of the $N_x$ initial substates, labeled, $n_x=1,...,N_x$, and terminates through one of the $M_x$ final substates, labeled $m_x = 1,...,$ $M_x$, for a reversible *on-off* connection KS (Fig. 2E), or $m_x = N_x +1-H_x,\ldots, N_x +M_x-H_x$, for an irreversible *on-off* connection KS (Fig. 1B), where $H_x$ (= $0,1,\ldots, N_x$) is the number of substates in state $x$ that are both initial and final ones. (In each of the states the labeling of the substates starts from 1). An event in state $x$ starts in substate $n_x$ with probability $W_{n_x}$. The first passage time PDF for exiting to substate $n_y$, conditional on starting in substate $n_x$ ($x \neq y$), is $f_{n_y n_x}(t)$, and $F_{n_x}(t) = \sum_{n_y} f_{n_y n_x}(t)$. Writing $f_{n_y n_x}(t)$ as, $f_{n_y n_x}(t) = \sum_{m_x} \omega_{n_y m_x} \tilde{f}_{m_x n_x}(t)$, emphasizes the role of the *on-off* connectivity, where $\omega_{n_y m_x}$ is the transition probability from substate $m_x$ to substate $n_y$, and $\tilde{f}_{m_x n_x}(t)\omega_{n_y m_x}$ is the first passage time PDF, conditional on starting in substate $n_x$, for exiting to substate $n_y$ through substate $m_x$.

***II.3. Relationships between the master equation and the path representation*** All the factors in Eq. (5) can be expressed in terms of the matrices of Eq. (1). $W_{n_x}$ and $f_{n_y n_x}(t)$ are related to the master equation by,

$$W_{n_x} = \left(\mathbf{V}_y \vec{\mathbf{P}}_y(ss)\right)_{n_x} / N_x,$$

and,

$$f_{n_y n_x}(t) = \left(\mathbf{V}_x \mathbf{G}_x(t)\right)_{n_y n_x}.$$

$f_{n_y n_x}(t)$ can be further rewritten as,

$$f_{n_y n_x}(t) = \sum_{m_x} \omega_{n_y m_x} \tilde{f}_{m_x n_x}(t),$$

and similarly for $\left(\mathbf{V}_x \mathbf{G}_x(t)\right)_{n_y n_x}$ we have,

$$\left(\mathbf{V}_x \mathbf{G}_x(t)\right)_{n_y n_x} = \sum_k \left(\mathbf{V}_x\right)_{n_y k} \left(\mathbf{G}_x(t)\right)_{k n_x}.$$

Note however that the factors in the right hand side in the above two sums are not equal but proportional,

$$\tilde{f}_{kn_x}(t) = \alpha_{x,k}(\mathbf{G}_x(t))_{kn_x} \quad ; \quad \alpha_{x,k} = -(\mathbf{K}_x)_{kk},$$

and

$$\omega_{n_y,k} = (\mathbf{V}_x)_{n_y,k} / \alpha_{x,k}.$$

### *III - RD Forms*

In this section we present the canonical forms of reduced dimensions and the relationships between the data, the canonical form topology, and the KS. Our canonical forms are based upon the path representation in section **II.2**.

**III.1. The rank of $\phi_{x,y}(t_1,t_2)$ and it's topological interpretation** For discrete time, $\phi_{x,y}(t_1,t_2)$ is a matrix, whose rank $R_{x,y}$ (= 1, 2, …), which is the number of non-zero eigenvalues (or singular values for a non square matrix) of it's decomposition, can be obtained without the need of finding the actual functional form of $\phi_{x,y}(t_1,t_2)$. Using Eq. (5), which gives $\phi_{x,y}(t_1,t_2)$ as sums of terms each of which is a product of a function of $t_1$ and a function of $t_2$, we can relate $R_{x,y}$ ($x \neq y$) to the topology of the underlying KS. When none of the terms in an external sum on Eq. (5), after the first or the second equality, are proportional, $R_{x,y} = \min(M_x, N_y)$ (Figs. 2A-2C, and Fig. 2F). Otherwise, $R_{x,y} < \min(M_x, N_y)$ (Fig. 2E, and Appendix B), and Eq. (5) is rewritten such that it has the *minimal* number of additives in the external summations,

$$\phi_{x,y}(t_1,t_2) = \sum_{n_y \in \{\tilde{N}_y\}} \left( \sum_{n_x=1}^{N_x} W_{n_x} f_{n_y n_x}(t_1) \right) F_{n_y}(t_2)$$

$$+ \sum_{m_x \in \{\tilde{M}_x\}} \left( \sum_{n_x=1}^{N_x} W_{n_x} \tilde{f}_{m_x n_x}(t_1) \right) \left( \sum_{n_y \notin \{\tilde{N}_y\}} \omega_{n_y m_x} F_{n_y}(t_2) \right). \quad (6)$$

This leads to the equality, $R_{x,y} = \tilde{N}_y + \tilde{M}_x$. $\tilde{N}_y$ and $\tilde{M}_x$ can be related to the KS's *on-off* connectivity. Consider a case where $M_x < N_y$, and there is a group of final substates in state *x*, $\{O_x\}$, with connections *only* to a group of initial substates in state *y*, $\{O_y\}$, and $O_x > O_y$ (see Fig. B4 in Appendix B). Then $\tilde{M}_x = M_x - O_x$ and $\tilde{N}_y = O_y$. (Further discussion and a generalization of this relationship are given in Appendix B).

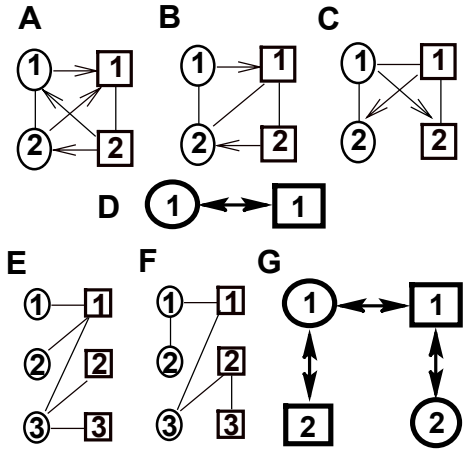

**FIG 2** Indistinguishable KSs. KSs **A-C** have the simplest RD form (**D**) of one substate in each of the states. KSs **A-C** are equivalent when they have the same $\phi_{on}(t)$ and $\phi_{off}(t)$. Equivalent KSs **E-F** have $R_{x,y}$=2, *x, y = on, off*, and tri-exponential $\phi_{on}(t)$ and $\phi_{off}(t)$. The corresponding RD form is shown in **G**.

***III.2. The RD form*** The $R_{x,y}$ s are obtained from the $\phi_{x,y}(t_1,t_2)$ s without the need of finding its' actual functional forms, thus constitute a fitting-free relationship between an ideal trajectory to the *on-off* connectivity and details of the underlying KS. Utilizing this relationship, the kinetic scheme space is divided into canonical forms,

RD forms, using the $R_{x,y}$s. Excluding KSs with symmetry, $R_{x,y}$ ($x \neq y$) is the number of substates in state *y* in the RD form. RD forms can represent underlying KSs with symmetry and irreversible connections because they are built from all four $R_{x,y}$s. The RD form has the minimal number of substates needed to reproduce the data. Connections in the RD form are only between substates of different states. For each connection in the RD form there is a WT-PDF that is not necessarily exponential. We denote by $\varphi_{x,ij}(t)$ the WT-PDF in the RD form connecting substates $j_x \rightarrow i_y$. $\varphi_{x,ij}(t)$ is a sum of exponentials with as many components as in $\phi_x(t)$.

***III.3. Mapping a KS into a RD form*** Mapping a KS into a RD form is based on clustering some of the initial substates in the KS into substates in the RD form, where initial substates in the KS that are not clustered are mapped to themselves. The KS's *on-off* connectivity determines whether an initial substate in the KS is clustered or mapped to itself. For a non-symmetric KS, initial substates in state *y* in the KS that contribute to $R_{x,y}$ ($x \neq y$) are mapped to themselves and those that do not contribute to $R_{x,y}$ are clustered, where initial-*y*-state substates in a cluster are all connected to the same final-*x*-state substate that contributes to $R_{x,y}$. (When substate $m_x$ has a single exit-connection to substate $n_y$, which is it's only entering-connection, substate $n_y$ is defined as the one contributing to the rank). For example, the KS in Fig. 2E is mapped into a RD form (Fig. 2G) when clustering substates $1_{off}$-$2_{off}$ into the RD form's substates $1_{off}$ where substate $3_{off}$ is mapped to itself giving rise to substate $2_{off}$ in the RD form, because only substate $3_{off}$ contributes to the rank $R_{on,off}$ among the *off* initial substates. A similar scenario is seen in the *on* state in this KS with a change of the labeling; substate $1_{on}$ is mapped to itself into substate $1_{on}$, and substates $2_{on}$-$3_{on}$ are clustered into RD forms substate $2_{on}$. Only substate $1_{on}$ contribute to $R_{off,on}$ among the *on* initial substates.

The clustering procedure determines the coefficients in the exponential expansion of the $\varphi_{x,ij}(t)s$. (Technical details for obtaining these WT-PDFs given a KS are discussed in Appendix B). Note that the clustering procedure, along with the fact that substates in the KS that are not initial or final ones do not affect the RD form's topology, reduce the KS dimensionality to that of the RD form.

***III.4. Examples and the utility of RD forms*** The simplest topology for a RD form has one substate in each of the states, namely, $R_{x,y} = 1$ ($x, y = on, off$), and the only possible choice for $\varphi_{x,11}(t)$ is $\phi_x(t)$ (Fig. 2D). This means that all the information in the data is contained in $\phi_{on}(t)$ and $\phi_{off}(t)$. Consequently, KSs with $R_{x,y} = 1$ ($x, y = on, off$) and the same $\phi_{on}(t)$ and $\phi_{off}(t)$ are indistinguishable (assuming no additional information on the mechanism is known). Examples of such KSs are shown in Fig. 2A-2C. The $R_{x,y}=1$ case was discussed in Refs. 28-30. The generalization of the equivalence of KSs for any case is straightforward using RD forms. KSs with the same $R_{x,y}$s and the same WT-PDFs for the connections in the RD form cannot be distinguished. Indistinguishable KSs with $R_{x,y} = 2$ ($x, y = on, off$) and tri-exponential $\phi_{on}(t)$ and $\phi_{off}(t)$ are shown in Figs. 2E-2F, and their corresponding RD form is shown in Fig. 2G.

Clearly, two KSs with different $R_{x,y}$s can be resolved by the analysis of a two-state trajectory. Among the advantageous of RD forms is by providing a powerful tool in resolving KSs with the same $R_{x,y}$s, and the same number of exponentials in $\phi_{on}(t)$ and $\phi_{off}(t)$, even without the need of performing actual calculations. This is done based only on distinct complexity of the WT-PDFs for the connections in the corresponding RD forms, e.g. compare the KSs in Fig. 3A and Fig. 3B, or, on different connectivity of RD forms, e.g. compare KSs in Figs. 3A-3B with the KS in Fig. 3C. The above means that it is impossible to find positive (> 0) transition rates for the KSs in Figs. 3A-3C that make the $\phi_{x,y}(t_1,t_2)$ s from these KSs the same, so these KSs can be distinguished by analyzing a two-state trajectory (excluding symmetric cases for which the $\phi_{x,y}(t_1,t_2)$ s factorizes to the product of $\phi_x(t_1)\phi_y(t_2)$ s).

Note that a RD form can preserve microscopic reversibility on the *on-off* level even when having irreversible connections. These can be 'balanced' by the existence of direction dependent WT-PDFs for the connections: microscopic reversibility in a RD form means that the $\phi_{x,y}(t_1,t_2)$ s obtained when reading the two-state trajectory in the forward direction are the same as the corresponding $\phi_{x,y}(t_1,t_2)$ s obtained when reading the trajectory backwards (as suggested in Ref. 39 for aggregated Markov chains). Using matrix notation, microscopic reversibility means, $\phi_{x,y}(t_1,t_2) = [\phi_{y,x}(t_1,t_2)]^T$, where $T$ stands for the transpose of a matrix.

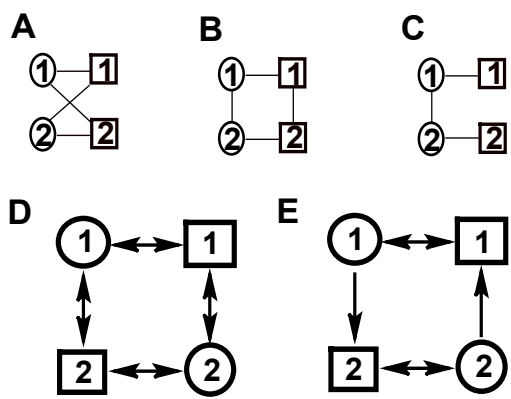

**FIG 3** Distinguishable KSs with $R_{x,y}$=2, *x, y* = *on, off* and bi-exponential and $\phi$ . (We exclude symmetry in this example). KS **3C** is distinct from KSs **3A** and **3B**, because the corresponding RD forms, **3E** and **3D**, respectively, have different connectivity. KS **3A** and KS **3B** are also distinct, because the WT-PDFs for the connections in the RD form of KS **3A** are exponential, whereas those of KS **3B** are direction-dependent and bi-exponentials.

The division of KSs into RD forms is useful also when, on top of the information extracted from the 'original' two-state trajectory, complementary details about the observed process are available. (Complementary details are obtained by analyzing different kind of measurements of the system, e.g. the crystal structure of the biopolymer, or by analyzing two-state trajectories while varying some parameters, e.g. the substrate concentration [13-15].) Suppose that the connectivity of the underlying KS is unchanged by the manipulation. Then, the additional information can be used to resolve KSs that correspond to the RD form found from the statistical analysis of the 'original' two-state trajectory, whereas any KS with a different RD form is irrelevant. Alternatively, when manipulating the system leads to a change in the connectivity of the underlying KS, or even to the addition or removal of substates, the RD forms obtained from the different data sets are distinct. RD forms can handle both scenarios; in the first case an adequate parameter tuning relates the RD forms obtained from the various sources, whereas in the second case the RD forms cannot be related by a parameter tuning.

*IV - Constructing the RD form from the data*

The RD form is built from finite data following a three-stage algorithm: (*1*) Obtain the spectrum of the $\phi_x(t)$s by the Padé approximation [56]. The spectrum of the WT-PDFs for the *x* to *y* connections in the RD form is the same spectrum as that of $\phi_x(t)$, because substates of the same state in the RD form are not connected. Differences lay in the pre-exponential coefficients. (*2*) Find the ranks of the $\phi_{x,y}(t_1, t_2)$s, and use it to build the RD form topology. (*3*) Apply a maximum likelihood procedure for finding the pre-exponential coefficients of the $\varphi_{x,ij}(t)$s. Our routine for maximizing the likelihood function uses its analytical derivatives.

We use the above three-stage algorithm in the construction of the RD form in Fig. 1C from finite data. The KS is shown in Fig. 1B. A two-state trajectory (Fig. 1A) is generated by simulating a random walk in the RD form. This is done by a modified Gillespie Mote-Carlo method. Each transition in the simulation happens in two steps. Assume the process starts at substate $i_x$. The first step chooses the destination of the next location, determined by the weights of making a transition $i_x \rightarrow j_y$: $w_{j_y i_x} = \overline{\varphi}_{x,j_y i_x}(0) / \sum_{j'_y} \overline{\varphi}_{x,j'_y i_x}(0)$. (Note that from the analytical expressions of Appendix B, the sum $\sum_{j'_y} \overline{\varphi}_{x,j'_y i_x}(0)$ is unity, but due to numerical issues it can be smaller than unity). The second step uses the particular $j_y$, and draws a random time out of a normalized density $\varphi_{x,j_y i_x}(t) / \overline{\varphi}_{x,j_y i_x}(0)$. The procedure is then repeated at the new location. It is much faster to generate a two-state trajectory using the RD form rather than simulating a random walk in the underlying KS.

Figure 4 displays the analytical and the experimental $\phi_{on}(t)$ and $\phi_{off}(t)$. The simulated data contained $10^6$ *on-off* events. The $\phi_x(t)$s are accurately obtained from the data for times such that their amplitudes are 2 orders of magnitudes smaller than its' maximal values. The Padé approximation method gives the correct amplitudes, rates, and number of components for these WT-PDFs.

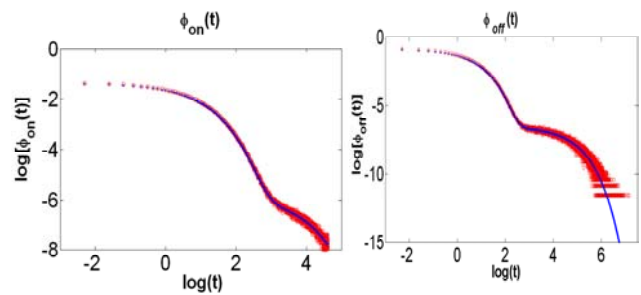

**FIG 4** The WT-PDFs of the *on* (left) and the *off* (right) states on a log-log scale. Shown are both the WT-PDFs obtained from a numerical solution of Eq. (2), full line, and by constructing these PDFs from a $10^6$ *on-off* event trajectory, circled symbols.

The next stage in the construction of RD form estimates the ranks of the $\phi_{x,y}(t_1,t_2)$ s. The correct values for the $R_{x,y}$ s, $R_{x,y} = 2 \ \forall x,y = on, off$, are easily obtained from the analytically $\phi_{x,y}(t_1,t_2)$ s by applying singular value decomposition (SVD), but the ratio of the large to small singular value is large ($\sim 10^3$). This result means that the contribution from the large singular value contains most of the signal, and corresponds to the limit of an infinitely long trajectory. Thus, one may expect technical difficulties in detecting the exact number of nonzero singular values from an experimental matrix, due to the limited number of events in the trajectory. To deal with this issue, we build a series of cumulative 2D WT-PDFs. The first order cumulative of $\phi_{x,y}(t_1,t_2)$ is defined by,

$$c_1\phi_{x,y}(T_1,T_2) = \int_0^{T_1} \int_0^{T_2} dt_1 dt_2 \phi_{x,y}(t_1,t_2),$$

and the generalization to higher order cumulative PDFs naturally follows,

$$c_n\phi_{x,y}(T_1,T_2) = \int_0^{T_1} \int_0^{T_2} dt_1 dt_2 c_{n-1}\phi_{x,y}(t_1,t_2).$$

A cumulative two dimensional PDF reduces the noise in the *original* PDF, but also preserves the rank of the original PDF. For each two dimensional PDF we obtain its spectrum of singular values and plot the ratio of successive singular values as a function of the order of the large singular value in the ratio. This plot should show large values for signal ratios, and a constant behavior with a value of about a unity for noise ratios. Figures 5A-5B shows the singular value ratio method applied on $\phi_{on,off}(t_1,t_2)$ and its first three cumulative PDFs. Both the second and the third cumulative PDFs show large values for the first two ratios and a constant behavior for larger ratios. This is a signature for a rank two histogram. (Note that this behavior is not seen in the original PDF and it's first cumulative.) For comparison, Fig. 5C plots the same quantities calculated for $\phi_{off,on}(t_1,t_2)$, all of which show the same constant behavior for ratios greater than one, indicating a rank one histogram. A rank one behavior is observed also for $\phi_{on,on}(t_1,t_2)$ and $\phi_{off,off}(t_1,t_2)$ (data not shown).

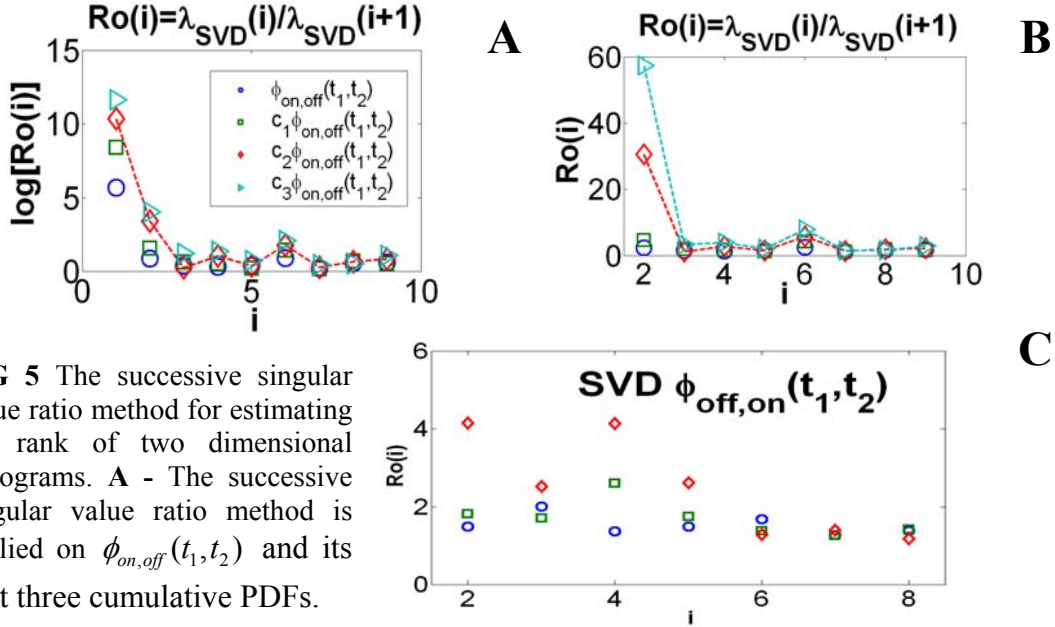

**FIG 5** The successive singular value ratio method for estimating the rank of two dimensional histograms. **A** - The successive singular value ratio method is applied on $\phi_{on,off}(t_1,t_2)$ and its first three cumulative PDFs. The first ratio contains most of the signal in all PDFs. **B** – The second and third order cumulative PDFs indicate a rank 2 PDF. **C** – The same method is applied on $\phi_{off,on}(t_1,t_2)$, and indicates a rank one PDF. Here the diamond and square symbols correspond to the second and first cumulative PDFs respectively, and the circled symbols correspond to the original PDF.

Based on these findings, a low resolution RD form with one *on* substate and two *off* substates is built. A search for 8 parameters, denoted by $\Theta = \{\alpha_{x,jHi}\}$, in the bi-exponential expansions of the $\varphi_{x,ij}(t)$s,

$$\varphi_{x,ji}(t) = \sum_{H=1}^{2} \alpha_{x,jHi} e^{-\lambda_{x,H} t}, \tag{7}$$

is performed by a maximum likelihood method with constraints. (Note that some of the $\alpha_{x,jHi}$s are zeros, but this information is not known a priori.) The likelihood function is given by,

$$L(\Theta | data) = \sum_{x,y} \sum_{i} \log(\phi_{x,y}(t_i, t_{1+i})).$$

The constraints in the algorithm demand that Eqs. (7) are positive for every value of $t$, and also the normalization of the WT-PDFs for the connection,

$$\sum_j \int_0^\infty \varphi_{x,ji}(t)dt = 1,$$

for every $x$ and $i$. The analytical derivatives of the likelihood function are used in the maximization procedure. The maximization is performed using the command '*fmincon*' in the software *Matlab*. (In fact, we minimize, $-L(\Theta\,|\,data)$.) The optimization procedure is performed in less than a minute for data made of $10^3$ events. In this case, a straightforward optimization gave the correct answer. (Note that the optimization can also find local maximum, so, for a general case, it is required to choose various sets of initial conditions.) For the studied case, the error bars for the elements in $\{\Theta\}$ are always less than a percent of the found values. (The error bars are found by inverting the Hessian matrix of second derivatives of the likelihood function with respect to the unknowns and substituting the solution for the unknowns. The diagonal elements of the obtained matrix give the variance of the fit [57].)

## V - Concluding remarks

This paper utilized the information content in a two-state trajectory for an efficient elucidation of a unique mechanism that can generate it. The KS space is partitioned into canonical forms, RD forms, that are (usually) not Markovian. The topology of a canonical form is determined by the ranks $R_{x,y}$s of the $\phi_{x,y}(t_1,t_2)$s. A RD form has connections only between substates of different states, which are usually non-exponential WT-PDFs. The relationship between the ranks $R_{x,y}$s and the KS's *on-off* connectivity is given. This relationship enables mapping a KS into a RD form based only on the *on-off* connectivity of the KS. The relationships between the ranks $R_{x,y}$s and the KS's *on-off* connectivity are based on path representation of the $\phi_{x,y}(t_1,t_2)$ s given in Eqs. (5)-(6). Among the advantages of RD forms is by constituting a powerful tool in discriminating among *on-off* KSs.

An example that builds a RD form from a data of $10^6$ *on-off* cycles shows the applicability of RD forms in analyzing realistic data. We thus believe that RD forms will be found useful in the analysis of single molecule measurements that result in two-state trajectories.

**APPEDIX A**

In this Appendix, we give expressions for $\phi_x(t)$ and $\phi_{x,x}(t_1,t_2)$ using the path representation. Here, and in Appendix B, $x \neq y$, unless otherwise is explicitly indicated. The expression for $\phi_x(t)$ is obtained from Eq. (5) by integrating over one time argument, $\phi_x(t) = \int_0^\infty \phi_{x,y}(t,\tau)d\tau = \int_0^\infty \phi_{y,x}(\tau,t)d\tau$, which leads to,

$$\phi_x(t) = \sum_{n_y=1}^{N_y}\sum_{n_x=1}^{N_x} W_{n_x} f_{n_y n_x}(t) = \sum_{m_x \in \{M_x\}} \sum_{n_x=1}^{N_x}\sum_{n_y=1}^{N_y} W_{n_x} \widetilde{f}_{m_x n_x}(t)\omega_{n_y m_x}. \qquad (A1)$$

The expression for $\phi_{x,x}(t_1,t_2)$ is obtained from Eq. (5) when introducing an additional summation that represents the random walk in state *y* that takes place in between the two measured events in state *x*,

$$\phi_{x,x}(t_1,t_2) = \sum_{n'_x=1}^{N_x}\sum_{n_y=1}^{N_y}\sum_{n_x=1}^{N_x} W_{n_x} f_{n_y n_x}(t_1) p_{n'_x n_y} F_{n'_x}(t_2)$$

$$= \sum_{m_x \in \{M_x\}} \sum_{n'_x=1}^{N_x}\sum_{n_y=1}^{N_y}\sum_{n_x=1}^{N_x} W_{n_x} \widetilde{f}_{m_x n_x}(t_1)\omega_{n_y m_x} p_{n'_x n_y} F_{n'_x}(t_2). \qquad (A2)$$

Here, $p_{n'_x n_y}$ is the probability that an event that starts at substate $n_y$ exits to substate $n'_x$, and is given by $p_{n'_x n_y} = \bar{f}_{n'_x n_y}(0)$, where $\bar{g}(s) = \int_0^\infty g(t)e^{-st}dt$ is the Laplace transform of $g(t)$.

Note that higher order successive WT-PDFs e.g. $\phi_{x,y,z}(t_1,t_2,t_3)$, do not contain additional information on top of the $\phi_{x,y}(t_1,t_2)$ s. When the underlying KS has no symmetry (i.e. the spectrum of $\phi_x(t)$, $x$ = on, off, is non-degenerate) and/or irreversible connections, it is sufficient to use $\phi_{x,y}(t_1,t_2)$ for $x \neq y$, where for other cases, $\phi_{x,x}(t_1,t_2)$ s, $x$ = on, off, contain complementary information.

**APPEDIX B**

In this Appendix, we give expressions for the WT-PDFs for the connections in the RD form, denoted by $\varphi_{x,ij}(t)$ s, for any KS. We do not consider symmetric KSs separately, because symmetry is apparent in the functional form of the $\varphi_{x,ij}(t)$ s. Further discussion regarding the topological interpretation of $\widetilde{M}_x$ and $\widetilde{N}_y$ in Eq. (6) is also given.

The waiting time PDFs for the connections in the RD form are uniquely determined by the clustering procedure in the mapping of a KS into a RD form. The clustering procedure is based upon the identification of substates in the *on-off* connectivity that contribute to the ranks $R_{x,y}$. The four ranks $R_{x,y}$s determine the RD form topology, and the mapping determine the incoming flux and outgoing flux for each substate in the RD form. This makes the RD forms legitimate canonical forms that preserve all the information contained in the two-state trajectory.

The technical details to obtain the $\varphi_{x,ij}(t)$s, given a KS, are spelled out below when considering two cases: (*1*) None of the terms in an external sum in Eq. (5), after the first or second equality, are proportional to each other, and (*2*) Some of the terms in an external sum in Eq. (5), after the first or second equality, are proportional to each other.

(*1.1*) *Reversible on-off connection KSs* Say, $M_x \geq N_y$, or equivalently $N_x \geq M_y$. (Fig. B1A with *x=off*). Based on the clustering procedure, there are $N_y$ substates in each of the states in the RD form, and as many as $2N_y^2$ WT-PDFs for the connections in the RD form. Initial substates in state *x* are clustered, and the expression for $\varphi_{x,n_y i_x}(t)$ reads,

$$\varphi_{x,n_y i_x}(t) = \frac{1}{N_{x,m_y}} \sum_{n_x} P_{y,m_y}(ss)(\mathbf{V_y})_{n_x m_y} f_{n_y n_x}(t). \tag{B1}$$

In Eq. (B1), we use the normalization $N_{x,m_y}$, defined through the equations,

$$N_x = \vec{\mathbf{1}}_x^T \mathbf{V_y} \vec{\mathbf{P}}_y(ss) = \sum_{m_y,n_x} P_{y,m_y}(ss)(\mathbf{V_y})_{n_x m_y} = \sum_{m_y} N_{x,m_y} = \sum_{n_x} N_{x,n_x}.$$

As notation is concerned, we set in Eq. (B1) $j_y \to n_y$ because there are $n_y = 1,...,N_y$ substates in state *y* in the RD form, and we can also employ the meaning of $n_y$ as the initial substates in state *y* in the underlying KS. Additionally, we associate $m_y$ on the right hand side (RHS), which has the meaning of final substates in the underlying KS, with $i_x$ on the left hand side (LHS), i.e. $m_y \to i_x$. Note that for a KS with only reversible *on-off* connections, $m_y = 1,...,M_y$, so the values of $m_y$ and $i_x$ can be the same.

The expression for $\varphi_{y,i_x n_y}(t)$ is different than that for $\varphi_{x,n_y i_x}(t)$ in both the normalization used and the factors that are summed, which is a result of the mapping of the initial substates in state $y$ to themselves. $\varphi_{y,i_x n_y}(t)$ is given by,

$$\varphi_{y,i_x n_y}(t) = \frac{1}{N_{y,n_y}} \sum_{m_x} P_{x,m_x}(ss)(\mathbf{V_x})_{n_y m_x} \tilde{f}_{m_y n_y}(t) \tilde{\omega}_{m_y} \quad ; \quad \tilde{\omega}_{m_y} = \sum_{n_x} \omega_{n_x m_y}. \quad (B2)$$

Note that here, $\varphi_{y,i_x n_y}(t) = \tilde{f}_{m_y n_y}(t) \tilde{\omega}_{m_y} = (\mathbf{G_y}(t))_{m_y n_y} \sum_{n_x} (\mathbf{V_y})_{n_x m_y}$. In Eq. (B2), we associate $m_y$ on the RHS with $i_x$ on the LHS, i.e. $m_y \to i_x$. Again, for a KS with only reversible transitions, the $i_x$s can have the same values as of the $m_y$s.

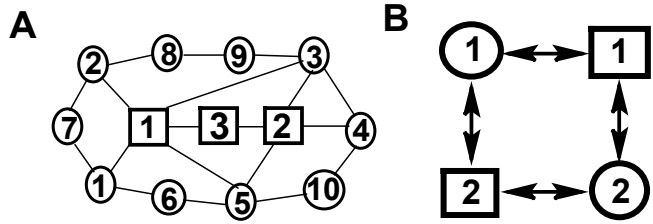

**FIG B1 A** – A reversible connection KS, with $N_{on} = M_{on} = 2$ and $N_{off} = M_{off} = 5$. **B** - The RD form of KS **A**. The RD form's substate $1_{off}$ corresponds to the cluster of the KS's *off* substates $1_{off}$-$3_{off}$ and $5_{off}$, because these are connected to substate $1_{on}$ in the KS, which contributes to the rank $R_{on,off}$. The RD form's substate $2_{off}$ corresponds to the cluster of the KS's *off* substates $3_{off}$-$5_{off}$, because these are connected to substate $2_{on}$ in the KS, which contributes to the rank $R_{on,off}$. Note that a particular initial substate can appear in more than a single cluster, which simply means that the overall steady-state flux into the substate in divided into several contributions. The initial *on* substates in the KS both contribute to $R_{off,on}$ so they are mapped to themselves in the RD form. The WT-PDFs for the connections can be obtained from Eqs. (B1)-(B2).

(*1.2*) *Irreversible on-off connection KSs* Obtaining the $\varphi_{x,ij}(t)$s for irreversible *on-off* connection KSs is similar to getting these WT-PDFs for reversible *on-off* connection KSs. The reason is that the clustering procedure is based on the *directional* connections between final substates in state $x$ and initial substates in state $y$. However, some technical details may differ. We consider two cases.

(*1.2.1*) Let $M_x \geq N_y$ and $M_y \geq N_x$. (Fig. B2A-B2B). Then, the WT-PDFs for the connections are given by,

$$\varphi_{x,n_y n_x}(t) = \frac{1}{N_{x,n_x}} \sum_{m_y} P_{y,m_y}(ss)(\mathbf{V_y})_{n_x m_y} f_{n_y n_x}(t) = f_{n_y n_x}(t), \qquad (B3)$$

and,

$$\varphi_{x,n_x n_y}(t) = \frac{1}{N_{y,n_y}} \sum_{m_x} P_{x,m_x}(ss)(\mathbf{V_x})_{n_y m_x} f_{n_x n_y}(t) = f_{n_x n_y}(t). \qquad (B4)$$

Note that for this case any $\varphi_{z,ij}(t)$ equal to the corresponding $f_{ij}(t)$. This is an outcome of the KS's topology for which in both the *on* to *off* and the *off* to *on* connections, the number of initial substates in a given state is lower than the number of final substates in the other state.

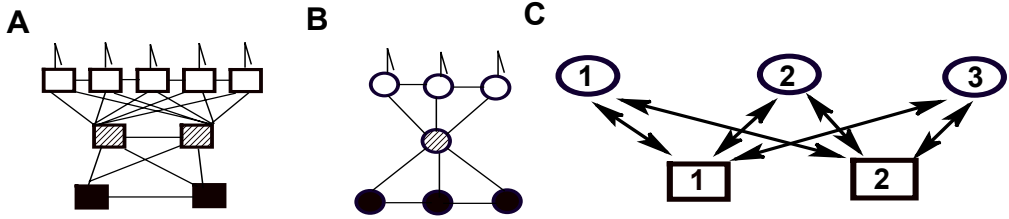

**FIG B2** An example for a KS with irreversible *on-off* connections, and $N_{on}=2$, $M_{on}=5$, $N_{off}=3$, and $M_{off}=3$. The KS is divided into two parts shown on **A** (*on* state) and **B** (*off* state) for a convenient illustration. The RD form is shown on **C**. The WT-PDFs for the connections can be obtained from Eqs. (B3) - (B4).

(*1.2.2*) Let $N_x > M_y$ and $N_y > M_x$. (Fig. B3A-B3B). Then, the WT-PDFs for the connections are given by,

$$\varphi_{x,j_y i_x}(t) = \frac{1}{N_{x,m_y}} \sum_{n_x} P_{y,m_y}(ss)(\mathbf{V_y})_{n_x m_y} \tilde{f}_{m_x n_x}(t) \tilde{\omega}_{m_x}, \qquad (B5)$$

and,

$$\varphi_{y,i_x j_y}(t) = \frac{1}{N_{y,m_x}} \sum_{n_y} P_{x,m_x}(ss)(\mathbf{V_x})_{n_y m_x} \tilde{f}_{m_y n_y}(t) \tilde{\omega}_{m_y}. \qquad (B6)$$

In Eqs. (B5)-(B6), we use the mapping $m_y \to i_x$ and $m_x \to j_y$ between the RHS and the LHS indexes. (In particular, $m_y - (N_y - H_y) = i_x$, and $m_x - (N_x - H_x) = j_y$).

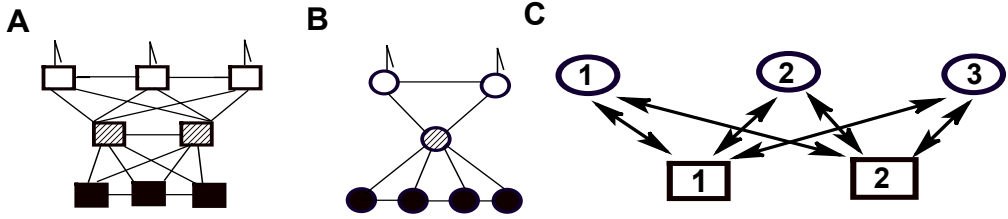

**FIG B3** An irreversible *on-off* connection KS with $N_{on}=3$, $M_{on}=3$, $N_{off}=4$, and $M_{off}=2$. The panels are divided as in Fig. B2. The WT-PDFs for the connections can be obtained from Eqs. (B5) - (B6).

(*2*) We turn now to deal with cases in which some of the terms in Eq. (5) are proportional, and therefore Eq.(6) is used for expressing $\phi_{x,y}(t_1,t_2)$. We consider only KSs with reversible *on-off* connections, but the same analysis is releavnt for KSs with irreversible *on-off* connections.

Let $M_x \leq N_y$, or equivalently $N_x \leq M_y$. (See Fig. B4A with $x=off$). So it follows that, $R_{x,y} < M_x$, which is a result of a special *on-off* connectivity. In particular, let $\{O_y\}$ and $\{O_x\}$ be the groups of substates in states *y* and *x* respectively, such that the substates in $\{O_x\}$ are connected only to the substates in $\{O_y\}$, and $O_y < O_x$. (In Fig. B4A, the group $\{O_{off}\}$ contains the substates $1_{off}$, $2_{off}$ and $3_{off}$, and the group $\{O_{on}\}$ contains the substates $1_{on}$ and $2_{on}$). Thus, both initial and final substates contribute to the rank $R_{z,z'}$ for $z \neq z'$, and the expressions for the $\varphi_{z,ij}(t)$ s are distinct in each of the following three regimes:

(*a*) For $n_x \notin \{O_x\}$ and $n_y \notin \{O_y\}$,

$$\varphi_{x,j_y i_x}(t) = \frac{1}{N_{x,n_x}} \sum_{m_y} P_{y,m_y}(ss) (\mathbf{V_y})_{n_x m_y} \tilde{f}_{m_x n_x}(t) \sum_{n_y \notin \{O_y\}} \omega_{n_y m_x}, \qquad (B7)$$

and,

$$\varphi_{y,i_x j_y}(t) = \frac{1}{N_{y \in O_y, m_x}} \sum_{n_y} P_{x,m_x}(ss)(\mathbf{V}_x)_{n_y m_x} f_{n_x n_y}(t), \tag{B8}$$

where $N_{y \in O_y, m_x} = \sum_{n_y \in \{O_y\}} P_{x,m_x}(ss)(\mathbf{V}_x)_{n_y m_x}$, and we associate $n_x \to i_x$ and $m_x \to j_y$.

(b) For $n_x \notin \{O_x\}$ and $n_y \in \{O_y\}$,

$$\varphi_{x,j_y i_x}(t) = \frac{1}{N_{x,n_x}} \sum_{m_y} P_{y,m_y}(ss)(\mathbf{V}_y)_{n_x m_y} f_{n_y n_x}(t), \tag{B9}$$

and,

$$\varphi_{y,i_x j_y}(t) = \frac{1}{N_{y,n_y}} \sum_{m_x} P_{x,m_x}(ss)(\mathbf{V}_x)_{n_y m_x} f_{n_x n_y}(t), \tag{B10}$$

where we associate $n_y \to j_y$ and $n_x \to i_x$.

(c) For $n_x \in \{O_x\}$ and $n_y \in \{O_y\}$,

$$\varphi_{y,i_x j_y}(t) = \frac{1}{N_{y,n_y}} \sum_{m_x} P_{x,m_x}(ss)(\mathbf{V}_x)_{n_y m_x} \tilde{f}_{m_y n_y}(t) \sum_{n_x \in \{O_x\}} \omega_{n_x m_y}, \tag{B11}$$

and,

$$\varphi_{x,j_y i_x}(t) = \frac{1}{N_{x \in O_x, m_y}} \sum_{n_x \in \{O_x\}} P_{y,m_y}(ss)(\mathbf{V}_y)_{n_x m_y} f_{n_y n_x}(t), \tag{B12}$$

where we associate $n_y \to j_y$ and $m_y \to i_x$.

Now, we use $O_y$ and $O_x$ for expressing $R_{x,y}$. When $M_x < N_y$ and $\{O_x\}$ and $\{O_y\}$ are as defined above,

$$R_{x,y} = M_x - (O_x - O_y). \tag{B13}$$

This result can be generalized to the case of $J$ groups in the underlying KS that are connected in the way defined above for the case of a single pair of groups. The generalized result reads,

$$R_{x,y} = M_x - \sum_j (O_{x,j} - O_{y,j}). \tag{B14}$$

These expressions imply that $\tilde{M}_x$ and $\tilde{N}_y$ on Eq. (6) are related to the KS's topology by,

$$\tilde{M}_x = M_x - \sum_j O_{x,j}, \tag{B15}$$

and,

$$\tilde{N}_y = \sum_j O_{y,j}. \tag{B16}$$

When $M_x > N_y$, and there are groups $\{Z_x\}$ and $\{Z_y\}$, with $Z_x < Z_y$, such that substates in $\{Z_y\}$ are connected only to substates in $\{Z_x\}$, we define $O_x = M_x - Z_x$ and $O_y = N_y - Z_y$, and Eq. (B13) holds. For $J$ such groups, we define $O_{x,j} = M_x / J - Z_{x,j}$ and $O_{y,j} = N_y / J - Z_{y,j}$, and Eqs. (B14)-(B16) hold.

For a KS with symmetry, $\tilde{M}_x$ and $\tilde{N}_y$ are chosen in a different way than the one relies on the *on-off* connectivity; for such a case, the choice that makes the number of additives in the external sums of Eq. (6) minimal simply groups the identical PDFs. The topology of the RD form is determined by the largest $R_{x,y}$.

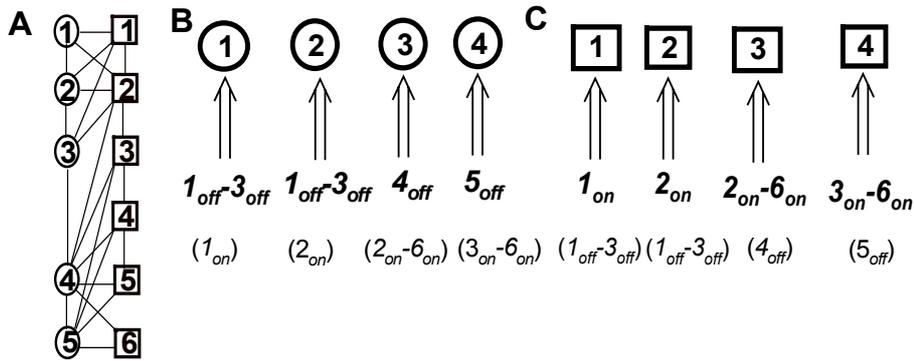

**FIG B4 A** A reversible connection KS with $R_{x,y}=4$ ($x \neq y$). The RD form's topology is shown on **B-C**. The clustering procedure and the parent substates (in the parenthesis) are indicated at the base of the double arrows. For example, substate $1_{off}$ in the RD form corresponds to the cluster of initial-*off*-substates $1_{off}$-$3_{off}$ in the KS. These are connected to substate $1_{on}$ in the KS. The WT-PDFs for the connections in the RD form can be obtained from Eqs. (B7)-(B12).